# Controlled assembly of single colloidal crystals using electro-osmotic micro-pumps

Ran Niu,*[1] Erdal C. Oğuz,[2,3] Hannah Müller,[1] Alexander Reinmüller[1], Denis Botin[1], Hartmut Löwen[2] and Thomas Palberg,[1]

[1] Institute of Physics, exp. Soft Matter Group KOMET336, Johannes Gutenberg University, D-55099 Mainz, Germany

[2] Institute of Theoretical Physics II: Soft Matter, Heinrich-Heine-University, D-40225 Düsseldorf, Germany

[3] present address: School of Mechanical Engineering and The Sackler Center for Computational Molecular and Materials Science, Tel Aviv University, Tel Aviv 69978, Israel

Abstract:

We assemble charged colloidal spheres at deliberately chosen locations on a charged unstructured glass substrate utilizing ion exchange based electro-osmotic micro-pumps. Using microscopy, a simple scaling theory and Brownian Dynamics simulations, we systematically explore the control parameters of crystal assembly and the mechanisms through which they depend on the experimental boundary conditions. We demonstrate that crystal quality depends crucially on the assembly distance of the colloids. This is understood as resulting from the competition between inward transport by the electro-osmotic pump flow and the electro-phoretic outward motion of the colloids. Optimized conditions include substrates of low and colloids of large electro-kinetic mobility. Then a sorting of colloids by size is observed in binary mixtures with larger particles assembling closer to the ion exchanger beads. Moreover, mono-sized colloids form defect free single domain crystals which grow outside a colloid-free void with facetted inner crystal boundaries centred on the ion exchange particle. This works remarkably well, even with irregularly formed ion exchange resin splinters.

*Corresponding author: Ran Niu

Email: ranniu@uni-mainz.de

**Introduction**

Colloidal suspensions of spherical particles readily crystallize, once their density and interaction strength are sufficiently large [1-5]. Their crystallization kinetics and mechanisms have been investigated in great detail in experiment and theory [6, 7]. Since the pioneering work of Tang et al. [8], there has been a growing interest in obtaining large single crystals (2D or 3D) with pre-determined location, structure and orientation. This is motivated by their use as model systems for fundamental issues in condensed matter physics [9] and their potential use as photonic or phononic materials [10,11]. Most common fabrication techniques are evaporation-related and lead to large scale sheet structures [12]. Other approaches rely on macroscopic gradients of different kind [13-15], or shear [16-18] or electro-convection [19]. This allows some control of the orientation of crystals and often produces very large single crystals. Simultaneous precise control of both the orientation and location of individual crystals from bulk suspensions is more demanding. Use of suitable 2D substrates for hard sphere crystal growth has been extensively studied [20, 21]. Under micro-gravity conditions hard sphere crystal nucleation and growth was templated on seed crystals placed in the bulk of the host suspension [22]. In addition to topological templates exploiting excluded volume interactions, also light grids exploiting photo-phoresis were used to fabricate individual 2D crystals and quasi-crystals [23, 24]. Use of holographic tweezer trapped seed structures allowed precise localization and orientation in 3D [25].

A different approach is taken by the use of catalytic, thermo-phoretic and electro-osmotic (eo) pumps, which rely on *local* convection fields and avoid template fabrication or elaborate optics. This has recently received much attention for transport, assembly and delivery of microscopic cargo [25-36]. Most of these work at elevated electrolyte concentrations or even at physiological conditions. This is a major draw-back for assembling colloidal crystals in dilute suspensions of charged colloidal spheres [4, 6]. First experimental observations of eo-pump related assembly and crystallization effects at ion exchange resin beads (which work at low levels of ionic impurities) date back some forty years [37], but largely remained unexplained. Today, the underlying mechanisms and related flow properties of different pump types receive growing experimental and theoretical attention [31, 33, 36, 38, 39].

In the present paper, we utilize an ion exchange based eo-pump to assemble 2D colloidal crystals at pre-determined positions from suspensions with µmolar impurity cation concentration. In eo-flow dominated experiments, similar to other works with different pump mechanisms, crystals grown were of low quality, showing multiple domains, different orientations, pronounced straining and many defects [35,40]. Therefore, in the present paper, we go beyond the previous works in several important points.40,41 First, we now also consider the influence of the electrophoretic motion of colloids on the assembly distance, which becomes relevant at low pump flow conditions. In fact, we assume that this central quantity for crystal growth is controlled by the competition between friction of the colloids in the inward solvent flow and their outward motion in the local electrostatic field. Furthermore, we undertake a number of modification in the experimental conditions allowing a systematic study of their influence on assembly distance and the crystal quality. We propose a simple 2D scaling theory to capture the dependence of assembly distance on different experimental boundary conditions. The experimental observations compare well to the analytical model and are further supported by the Brownian Dynamics (BD) computer simulations.

## Methods

### Experimental

The central part of our experiments is literally taken by the ion exchanger particles. We used both spherical and irregular shaped species. Commercial micro-gel spheres are made of sulfonate functionalized cross-linked divinylbenzene with average diameters of 15 μm (CK10S, Mitsubishi Chemical Corporation, Japan), 45 μm and 67 μm (CGC50×8, Purolite Ltd, UK), denoted as IEX15, IEX45 and IEX67. In addition, small irregular shaped IEX splinters (lateral extension < 100 μm) were obtained from crushing 1-2 mm IEX resin spheres (Amberlite K306, Roth GmbH, Germany) in a mortar. To qualitatively check the exchange rate per mass, we performed conductivity measurements in standard electrolyte solutions (details are shown in the supporting information (SI)). We found the exchange rate per mass of the resin based IEX to be about one order of magnitude larger than for the micro-gel IEX. For the later we obtain an exchange rate of $5 \times 10^{-17}$ mol/s, decreasing with $t^{-1/2}$, as expected for diffusion-limited exchange. Comparing this rate to the available amount of exchangeable cations shows that an exhaustion of the cation storage can be expected to become noticeable on the time scale of about 1500 s. Full exhaustion will not be reached due to the replenishment of sodium by the substrate, but the pump performance can be expected to decrease slowly with time.

Polystyrene (PS) spheres with sulfonate functional groups and with different diameters were purchased as 10% w/v suspension (MicroParticles Berlin GmbH, Germany). Their properties are compiled in Table 1. Diameters were determined by the manufacturer using electron microscopy. The electrophoretic mobility μep of isolated colloidal spheres was measured in deionized water at contact with ambient air by micro-electrophoresis (details shown in the SI). We observe values ranging $-2.6 \leqslant \mu_{ep} \leqslant -2.0 \times 10^{-8}$ m$^2$ V$^{-1}$ s$^{-1}$ for colloids of sizes ranging 1 μm $\leqslant 2a \leqslant$ 20 μm. We used standard electro-kinetic theory [42], to calculate the corresponding ζ-potentials and effective charges Zeff assuming a Debye-Hückel-type electro-static potential (see also below Eqns. 9a and 9b). Results for Zeff as a function of particle diameter are shown in Fig. S1. Over the range of diameters Zeff shows a power-law dependence and the best least square fit yields $Z_{eff} \propto a^{1.8}$.

Suspensions used for experiments were prepared by diluting the stock solution with doubly distilled water and were thoroughly deionized in contact with mixed bed ion exchanger (Amberlite K306, Roth GmbH, Germany) for extended times.

Substrates were soda lime glass microscopy slides of hydrolytic class 3 (VWR International, Germany). The glass slides were washed with alkaline solution (Hellmanex® III, Hellma) by sonication for 30 min then rinsed with tap water, and finally washed several times with doubly distilled water. They were dried in a laminar flow box under dust free conditions. To investigate the influence of substrate charging on pump activity, some glass slides were dip-coated in 2% aqueous solutions of Dimethyloctadecyl[3-(trimethoxysilyl)propyl] ammonium chloride (DMOAP, Merck, Germany) for different times. After coating, the slides were rinsed with doubly distilled water and dried again. Prepared slides were kept at ambient temperature and humidity conditions for some time before use. The background ion concentration obtained in our geometry by sodium release from the slides upon contact with water was estimated from manufacturer data to be on the order of $10^{-8}$ mol/L. This value is supported by the observation that the addition of merely 0.5 μmol/L of NaCl considerably prolonged the pumping activity. Doppler velocimetry with PS tracers was used to determine the electro-osmotic mobility $\mu_{eo}$ in contact with CO$_2$-saturated water, and standard electro-kinetic theory was used to calculate the corresponding zeta potentials from the reduced mobility, $\mu^*$ [43]. For the uncoated slides, we found $\mu^*_{eo} = -6.9$ corresponding to a ζ-potential of -114 mV in accordance with manufacturer values for this soda lime glass. It drops continuously with increasing time of DMOAP treatment to reach -50 mV ($\mu^*_{eo} = -3.0$) after one and -20 mV ($\mu^*_{eo} = -1.2$) after three hours of coating, denoted as DMOAP-1 and DMOAP-3, respectively. Adsorption of positively charged DMOAP on the negatively charged glass slides thus lowers the substrate potential and care has to be taken to avoid crossing the iso-electric point. For coating times longer than 4 h the colloids were found to stick irreversibly to the glass and no mobility data could be obtained. In the experiments reported below, we therefore used uncoated substrates, DMOAP-1 and DMOAP-3.

**Table 1** Properties of polystyrene cargo particles showing lab code, diameter, manufacturer batch number, effective charge $Z_{eff}$ and electrophoretic mobilities.

| Lab code | PS2 | PS4 | PS7 | PS10 |
|---|---|---|---|---|
| Batch No. | PS-F-B256 | PS/Q-F-B1203 | PS/Q-F-L771 | PS/Q-F-B1278 |
| 2a (µm) | 2.39 ± 0.04 | 4.1 ± 0.11 | 7.6 ± 0.1 | 10.7 ± 0.1 |
| $Z_{eff}$ | $9.0 \times 10^4$ | $2.6 \times 10^5$ | $8.8 \times 10^5$ | $4.06 \times 10^5$ |
| $\mu_{ep}$ ((µm/s)/(V/cm)) | -2.2±0.2 | -2.1±0.2 | -2.6±0.35 | -2.5±0.3 |

Custom sample cells were made from a circular Perspex ring (height 1 mm, inner diameter 20 mm) attached to a microscopy slide by hydrolytically inert epoxy glue. The cells were filled with about 0.4 ml of thoroughly deionized colloidal suspension and quickly sealed by a second glass slide to avoid contamination with air-borne $CO_2$ and dust. Within minutes, the colloids sedimented forming a dilute monolayer at the cell bottom. The IEX particles were then individually dropped into the suspension filled cell and the cell resealed. The IEX particles settle rapidly onto the substrate without much lateral deviation from the dropping point. For this protocol, the residual ion concentration is estimated to be close to $(2-3) \times 10^{-7}$ mol/L with about $1 \times 10^{-8}$ mol/L provided by cations instantaneously released by the substrate. Some samples were left at contact with ambient air, and these are believed to be saturated with $CO_2$. In this case, the initial electrolyte concentration resulting from dissociated carbonic acid is estimated to be 5 µmol/L [44]. In the following all the $CO_2$-saturated conditions are denoted by $c_S = 5$ µmol/L to distinguish with the deionized condition. In both cases, electrolyte gradients are present quasi instantaneously and become stationary within minutes, as seen from spatially resolved photometric measurements with added pH-indicator solutions [45]. Pumping activity remains stationary for up to an hour under deionised conditions and up to 24 hours at 1 µmol/L of added 1:1 electrolyte, before it slowly decreases.

Samples were placed on the motorized stage (IM120x100, Märzhäuser, Wetzlar, Germany) of an inverted optical microscope (DMI4000 B, Leica, Wetzlar, Germany) equipped with a 5x or 10x magnifying objective. Images and videos (at frame rate of 1 s$^{-1}$) were recorded with standard consumer DSLRs. The positions of tracer particles were extracted from the videos using a home-written python script. The essential part of particle detection was done using the OpenCV function HoughCircle. Generally, the position of a circular particle is mathematically represented as $(x-x_{cent})^2 + (y-y_{cent})^2 = r^2$, where $(x_{cent}, y_{cent})$ is the center of the circle, and r is its radius. To obtain the particle position, the Hough Gradient Method was used to extract the edge from image gradient information. Then, the center of the circular edge was used to extract the location of the particle center, via the above equation. Positions of particles were further used to determine the center-center separation, $d_{CC}$. Shortest distances between particle edges were used to determine the projected surface separation, $d_S$.

**Colloid assembly using electro-osmotic pumping Colloid assembly using electro-osmotic pumping: Theory and Simulation**

General Scheme

It is instructive to give a general scheme of the electro-osmotic pumping situation before going into details of theoretical description and simulation. In Fig. 1, we schematically illustrate our present understanding of the IEX-based eo-pump used for colloid assembly. The IEX exchanges stored protons for residual cationic impurities. A diffusio-electric field E0 is built up by the different diffusivities of protons and the exchanged cations. Acting on the substrate double layer, it induces a converging eo-flow (large blue arrows) [41]. This field locally acts on a colloid particle inducing an outward electro-phoretic (ep) drift relative to the solvent (cf. green arrows). Due to the fluid incompressibility, the eo-flow is naturally three-dimensional. At low flow conditions, however, the vertical convection can be neglected, and thus we can consider the system to be effectively two-dimensional. Furthermore, colloidal particles assemble at some distances d away from the IEX, which is the $d_{CC}$ used in simulation and the $d_S$ used in experiment as labelled accordingly in Fig. 1.

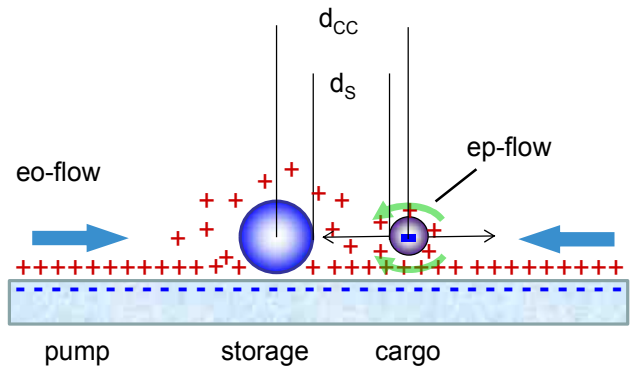

**Fig. 1** (color online): Sketch of eo-pumping (vertical cut) with a spherical IEX as storage, a substrate as pump, and an individual negatively charged colloid hovering at a height $z^*$ as cargo. The IEX released proton gradient induces an inward pointing diffusio-electric field, $E_0$, which in turn induces a radially converging electro-osmotic (eo) flow of the diffuse double-layer charge of the substrate. By friction, this flow drives the solvent towards the IEX (large light blue arrows). Due to friction, the cargo is carried along. Locally, the field, $E(r)$, induces an electro-phoretic (ep) solvent flow along the surface of the colloid (green arrows) leading to relative electro-phoretic motion of the colloid which is directed outward and opposes the inward solvent flow. The colloid is stalled at the assembly distance $d$, when the two opposing forces (thin arrows) cancel in the IEX frame. The two different measures of this central quantity, the projected centre-centre distance $d_{CC}$ and the projected surface-surface distance $d_S$, are also indicated.

Simple scaling theory

In our simple scaling theory, we assume a two-dimensional horizontal motion of the colloids at a constant vertical distance $z^*$ from the underlying substrate. For simplicity, we assume low flow conditions and neglect any vertical solvent flow. Then the height $z^*$ is approximately given by a vertical force balance

$$F_g = F_C \tag{1}$$

between the gravitational force $F_g$ and the repulsive force $F_C$ from the charged substrate. The gravitational force $F_g$ points downwards and is given by $F_g = \rho\, 4\pi a^3 g / 3$ where $g$ is the gravitational acceleration, $\rho$ the buoyant mass density of the colloids with respect to the solvent and $a$ the colloid radius. The wall repulsion scales as $F_C \propto \sigma\, Z_{eff} \exp(-\kappa z^*)$, where $\sigma$ is the effective surface charge density, $\kappa$ is the local inverse Debye screening length, and $Z_{eff}$ is the effective colloidal charge. Note that in the saturation limit of charge renormalization, the latter scales with the colloidal radius [46] as:

$$Z_{eff} \propto a(1+\kappa a) \tag{2}$$

From the vertical force balance we obtain a height $z^*$ scaling with $a$ as $-1/\kappa \ln(a^3 / Z_{eff}(a))$. The theoretical expectation for the proportionality constant in Eqn. (2) is $4/\lambda_B$, where $\lambda_B$ is the Bjerrum length, which in water amounts to 7 Å [47]. Electro-kinetic experiments at low colloid concentrations yield values around $2/\lambda_B$ in water and water/glycerol based systems and $1/\lambda_B$ in organic systems [48,49]. For the PS particles investigated here, we obtain $1.4/\lambda_B$. However, for the range of colloid sizes investigated, the dependence can also be fitted well by a simple power law yielding $Z_{eff} \propto a^{1.8}$ (Fig. S1).

We now consider the projected two-dimensional lateral motion of the colloids. Under conditions of low eo-pump flow, a single colloid basically experiences two opposing effects which we intend to model in a simple way. First of all, the colloid is dragged by the solvent flow field $u$ towards the IEX. This flow is assumed to be purely lateral exerting a Stokes drag force $F_{friction} = \gamma u$ on the colloids with $\gamma = 6\pi\eta a$ denoting Stokes drag coefficient. We assume for the solvent flow field that it is cylindrical symmetric around the IEX and scales as $u(r,z) \propto \mu_{eo} E_0 \exp(-\kappa z)/r$. The exponential dependence arises from electro-osmosis near the charged substrate of (material dependent) electro-osmotic mobility $\mu_{eo}$, while the $1/r$ dependence with the cylindrical distance comes from the incompressibility condition of the solvent in a cylindrical geometry.

The second, opposing force $F_{ep}$ stems from electrophoresis and is caused by the local electric field at the location of the colloid. A diffusio-electric field is built up throughout the electrolyte concentration gradient by the different diffusivities of protons and the exchanged ions. Most of the protons have diffused away establishing a radial symmetric effective stationary charge density profile. Under this condition, the electric field decays approximatively as $1/r^2$. Therefore the electro-phoretic force $F_{ep}$ acting on a single colloid is proportional to $\gamma\, \mu_{ep}/r^2$ where $\mu_{ep}$ is the electro-phoretic mobility of the colloids. The latter may in some cases depend on the colloidal diameter but no systematic dependence was observed for the present particles [49]. This dependence is therefore neglected for the parameters considered here. Then horizontal force balance

$$F_{friction} = F_{ep} \qquad (3)$$

sets the stationarity condition in lateral direction for a single colloid. This condition can equivalently also be expressed in terms of colloidal velocities *via* equating the solvent flow field $u = F_{friction}/\gamma$ to the electrophoretic velocity $v_{ep} = F_{ep}/\gamma$.

Inserting the scaling expressions from above into the lateral force equilibrium yields a scaling prediction for the stationary distance d* of the colloids from the IEX in terms of *a* as:

$$d^* \propto Z_{eff}(a)/\mu_{eo} a^3 \qquad (4)$$

Assuming no dependence of $\mu_{ep}$ and $\mu_{eo}$ on the colloidal radius *a*, we obtain a scaling of d* as $1/a$ or $1/a^2$ depending on whether the colloidal charge scaling is quadratic or linear in *a*. Therefore the simple theory predicts a scaling of d* as $1/a^\alpha$ with an effective exponent $\alpha$ between 1 and 2. Using the experimental data of Fig. S1 the exponent is expected to be close to $\alpha = 1.2$.

This simple treatment is expected to yield a qualitatively correct description in low flow situations [40]. It should break down for moderate and large flows, where solvent backflow and the 3D advection of cargo may not be neglected any longer and cargo is assembled close to the IEX [40,50] or even lifted above the substrate [27-36, 41, 45]. Measurements of the assembly radius of a given colloid species around a given IEX type on different substrates and of its size dependence for differently sized colloids can check these assumptions experimentally.

Brownian Dynamics simulations

Our simple view is also adapted for our two-dimensional Brownian dynamics (BD) computer simulations. These are aiming at verifying the idea that under competing flows the colloids can be stalled away from the substrate and that for sufficiently shallow potential minima, they can assemble according to their own interactions without the need to conform to an arbitrarily formed seed surface. We therefore consider the colloids to be trapped by the two radially symmetric opposing forces. First, the effective attractive trap force in 2D exerted by stationary a radial flow field can be written as an effective force as follows [40]:

$$\mathbf{F}_{friction}(\mathbf{r}) = \gamma \mathbf{u}(\mathbf{r}) = -\gamma \frac{A}{r^2}\mathbf{r} \qquad (5)$$

where **u** is the flow velocity at **r** (relative to the origin with r = |**r**|), and A is a positive flow amplitude. Herein, we physically assume that the convective flow is sufficiently weak that accumulated particles stay confined to the monolayer at $z = z^*$ due to gravity. The inverse distance dependency of the flow is again justified by the incompressibility of the fluid ($\nabla \bullet \mathbf{u}(\mathbf{r}) = 0$) in two dimensions and also found in simulations of the eo-pump flow in cells of small ehight (H = 1mm) [41]. Furthermore, following our simple scaling arguments, particles are exposed to an electro-phoretic force, $\mathbf{F}_{ep}$, caused by the quasi-stationary radial electrical field $\mathbf{E}(\mathbf{r})$:

$$\mathbf{F}_{ep}(\mathbf{r}) = B\mathbf{E}(\mathbf{r}) = \frac{B}{r^3}\mathbf{r} \qquad (6)$$

with an amplitude $B > 0$.

The equation for the trajectory $\mathbf{r}_j(t)$ of the colloidal particle $j$ undergoing Brownian motion (neglecting hydrodynamic interactions) in a time step $\delta t$ reads as:

$$\mathbf{r}_j(t+\delta t) = \mathbf{r}_j(t) + \frac{D_0}{k_B T}\mathbf{F}_j(t)\delta t + \mathbf{u}(\mathbf{r}_j)\delta t + \delta \mathbf{W}_j \qquad (7)$$

where $D_0 = k_B T/\gamma$ denotes the free colloid diffusion coefficient, and $\mathbf{F}_j(t)$ is the total conservative force acting on colloid $j$. Note the absence of any inertial term, which is justified by the low Reynolds number, $Re \ll 1$. The contributions to this force are stemming from the pair interactions between the colloids (i.e. $V(s)$, see below), the electro-phoretic force field, $\mathbf{F}_{ep}$ from Eq. (6), and a repulsive particle-boundary interaction due to an outer circular boundary. The latter is chosen to be a truncated and shifted 6-12 Lennard-Jones potential confining the particles within a disk of radius $R$. In some cases also an inner repulsive boundary is introduced to mimic the assembly conditions encountered at moderate flows. The third term on the right hand side of Eq. (7) is due to the solvent flow in which the cargo is transported (c.f. Eq. (5)). Finally, the random displacement $\delta \mathbf{W}_j$ is sampled from a Gaussian distribution with zero mean and variance $2D_0\delta t$ (for each Cartesian component) fixed by the fluctuation-dissipation relation.

The colloid-colloid interactions are modelled by a Yukawa pair potential

$$V(s) = V_0 \frac{\exp(-\kappa s)}{s} \qquad (8),$$

where $s$ is the colloid-colloid separation (centre-centre), $\kappa$ is the screening parameter, and $V_0$ denotes the interaction amplitude. For charged suspensions, the latter can be well described within the Debye-Hückel theory [50] yielding the following expressions

$$V_0 = \frac{Z_{eff}^2 e^2}{4\pi\varepsilon} \left( \frac{\exp(\kappa a)}{1+\kappa a} \right)^2 \qquad (9a),$$

and

$$\kappa \approx \sqrt{\frac{e^2}{\varepsilon k_B T}(n_{i,\infty} z_i^2 + n Z_{eff} z_i^2)} \qquad (9b).$$

Here, $z_i$ is the micro-ion valence assumed to be unity, $\varepsilon = \varepsilon_0 \varepsilon_r$ is the dielectric permittivity, and $n$ is the local colloid density. We implicitly assume that the case of high macro-ion densities is also captured by mapping the effects of counter-ion condensation [51] and macro-ion-shielding [52] on an effective macro-ion charge $Z_{eff}$, differing from the bare charge $Z$ valid at infinite dilution [53]. Note that Eq. (9b) accounts for screening by both the background number density of small electrolyte ions of species $i$ and the self screening of colloids by counter-ions.

**Results**

Experiments

Eo-flow velocities and assembly patterns differ considerably on different substrates. The largest flows are obtained using large splinters on uncoated substrates under thoroughly deionized conditions. The slowest flows are observed for long coated substrates, small spherical IEX particles and $CO_2$ saturation.

Interestingly, we observe an a priori unexpected flow strength dependence in the sorting of the colloids. At large flows, the smaller colloids assemble much closer to the IEX particles than the larger colloids. Figs. 2a and 2b give two examples for a mobile and an immobile IEX45 assembling a mixed cargo of PS7 and PS10, respectively. In Fig. 2a, a modular micro-swimmer is formed, and the whole complex shows directed motion with some μm/s. In Fig. 2b, IEX45 is pinned and the assembly remains at place. However, a pronounced convection cell builds. This is seen from the blurred images of PS7 lifted far above the focal plane located close to the uncoated substrate. Using large splinters on uncoated substrates even larger colloids are convected upward. Moreover, a pronounced influence of the splinter shape of the flow and the colloid assembly in the convection cell are observed (c.f. video 1).

On DMOAP-1 substrate and at $c_s = 5$ μmol/L, the flow is considerably reduced. Fig. 2c shows that now even PS10 stays settled onto the substrate. The assembly distances appear to be increased as compared to the previous cases, and they are similar for both colloidal species. However, they fluctuate strongly. All situations shown in Figs. 2a-c are not captured by our simple model. Fig. 2d was recorded under deionized conditions on DMOAP-3 using a small, compact splinter. Now, a stable assembly distance is observed as the inner boundary of a crystal of PS10. PS2 is assembled further outward. The sequence shown therefore documents a remarkable reversal of the assembly sequence and suggests to focus the systematic investigations of high quality crystal assembly on low flow situations with fixed IEX particles.

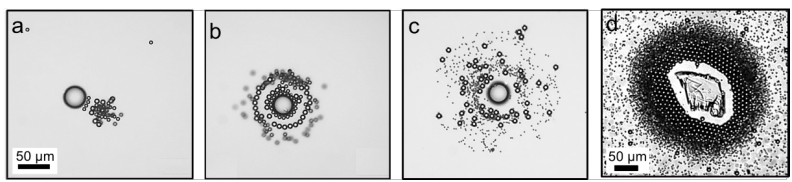

**Fig. 2** Optical micrographs of mixed cargo assembly. a) Swimmer formed of IEX45 and a binary mixture of PS7 and PS10 on uncoated substrate. b) eo-pump of immobilized IEX45 on uncoated substrate under deionized conditions and binary mixture of PS7 and PS10. In a) and b), PS7 is assembled closer to the storage particle. c) Binary mixture of PS2 and PS10 on DMOAP-1, $c_s$ = 5 µmol/L. No clear sorting occurs, and both cargo types stay sedimented. The scale bar of a) to c) is indicated in a). d) IEX splinter on DMOAP-3 under deionized conditions assembling a binary mixture of PS2 and PS10. A PS10 polycrystal is assembled inside a roughly circular region of PS2.

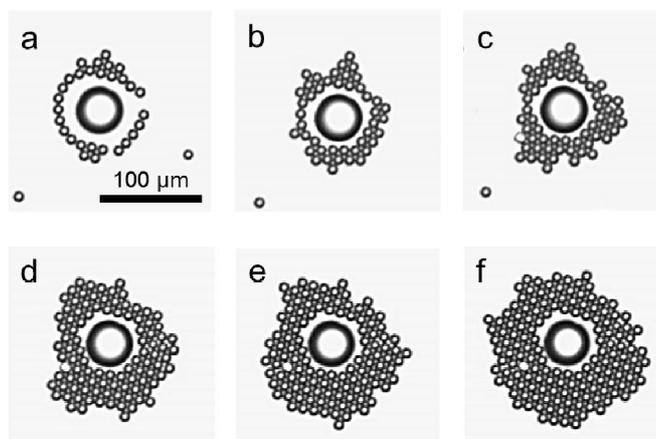

**Fig. 3** Typical stages of single crystal formation: PS10 at IEX45 on DMOAP-3 substrate, $c_s$ = 5 µmol/L. 100 µm scale bar applies to all images. a) Ring formation, b and c) addition of further rows; d and e) formation of inner facets and crystalline order, annealing of grain boundaries and defects; f) large single crystallite formed after annealing.

Under low flow condition (DMOAP-3 and $c_s$ = 5 µmol/L), cargo particles assemble to form a single crystal. The principle steps are shown in Figs. 3a-f. First a loose ring forms (a), then further layers are added to the ring (b and c). Crystalline order is typically formed, once three or more layers have formed after several minutes (d). The order increases in quality over time and both defects and grain boundaries anneal. During crystal formation and annealing, the inner boundary becomes facetted (e). Depending on the initial cargo density, a perfect single crystal forms within some 20-80 minutes (f). In 5-10% of experiments, a single grain boundary remains.

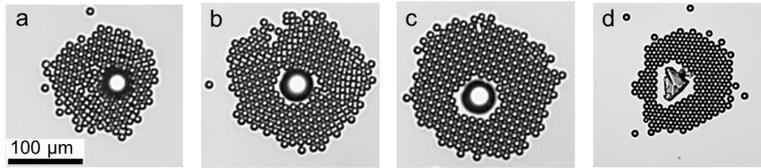

**Fig. 4** Assembly patterns of PS10 in eo-pumping with fixed IEX45: a) mono-layer cluster with no crystalline order on uncoated substrate, b) multi-domain crystal on DMOAP-1 and $c_S$ = 5 μmol/L and c) single crystal on DMOAP-3 and $c_S$ = 5 μmol/L. d) Single crystal formed by PS10 at IEX splinter on DMOAP-3 and $c_S$ = 5 μmol/L.

The assembly pattern of colloids under different flow conditions varies dramatically (Fig. 4), ranging from mono-layer cluster with no crystalline order (Fig. 4a), to a multi-domain crystal with many defects and dislocations (Fig. 4b), and even to a single defect-free crystal (Figs. 4c and 4d). Upon observing the crystal growth statistically at 150 IEX45, we find that most single crystals have a large assembly distance d at the stage of the ring formation. The average distance d shrinks with time, and shows a further strong variation along the facets upon the occurrence of facetted crystals (Fig. 3). The finite assembly distance allows for cargo rearrangements, so that the crystal structure commensurates with any of the IEX shape (c.f. Fig. 4d). This is confirmed by the quantitative results shown in Fig. 5, where three distinct regions of noncrystalline clusters (black), multi-domain crystals (red), and single crystals (blue) are sorted by the assembly distance *d*.

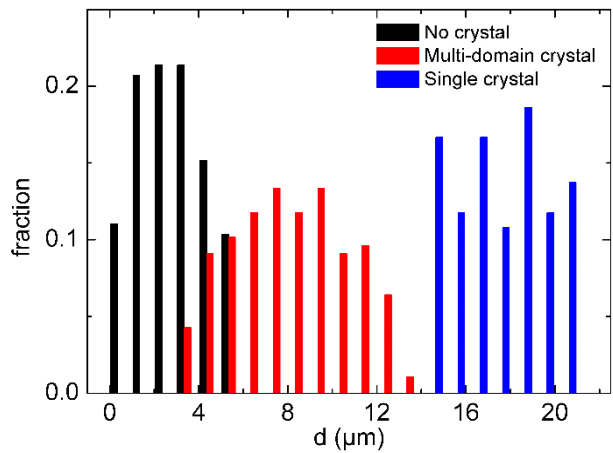

**Fig. 5** Histograms of the assembly patterns versus assembly distance (d) for the ring formation of PS10 in eo-pumping formed by an IEX45 under different flow conditions.

Nex, we investigate how the choice of experimental boundary conditions influences the assembly distance. We start with the dependence of final projected surface distances on the mobility ratio utilizing IEX45 and PS10. An increased DMOAP deposition time reduces the electro-osmotic mobility $\mu_{eo}$ along the substrate, while PS10 has a constant electro-phoretic mobility of $\mu_{ep}$ = -2.5 μms$^{-1}$/Vcm$^{-1}$. Note that the experimentally accessible range of $\mu_{eo}$ is limited. The upper limit of $\mu_{eo}$ in water suspension is set by material dependent surface charge (quartz is known to be the highest). One could further lower $\mu_{eo}$ by elongating the deposition time, however, at deposition time longer than 4 h, cargo sticks to the substrate. In Fig. 6, we observe a roughly linear increase of $d$ with deposition time. This corresponds to a roughly linear decrease with increasing mobility ratio, as is expected from Eqn. (4). In fact, the best fit to the data of Fig. 6 returns a slope of (0.89±0.06).

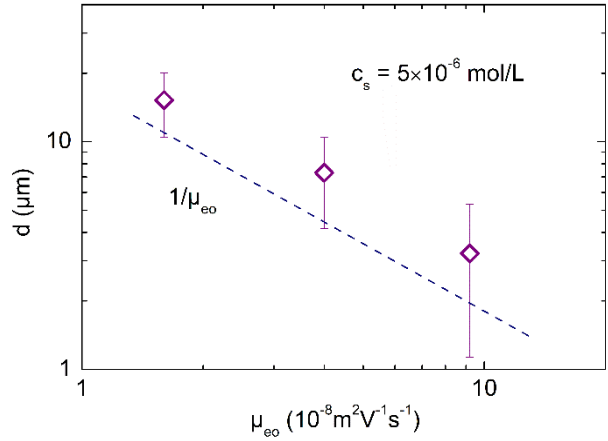

**Fig. 6 (color online)** Double logarithmic plot of the assembly distance $d$ of PS10 at spherical IEX45 on DMOAP-3, DMOAP-1 and uncoated substrates ($c_S$ = 5 μmol/L) resulting in an increasing $\mu_{eo}$. For comparison we show the theoretical expectation from Eqn. (4) as straight dashed line.

The influence of background salt on the assembly distance was qualitatively checked by changing the $CO_2$ saturation. Compared with deionized condition, $CO_2$ saturated condition only changes the mobility of surfaces (both substrate and cargo) without changing the exchange rate of the IEX. Without knowing how much μeo and μep are changed, Fig. 7 shows that $CO_2$ saturation increases the assembly distance of PS10 at IEX45 on the same substrates (from top to the bottom row).

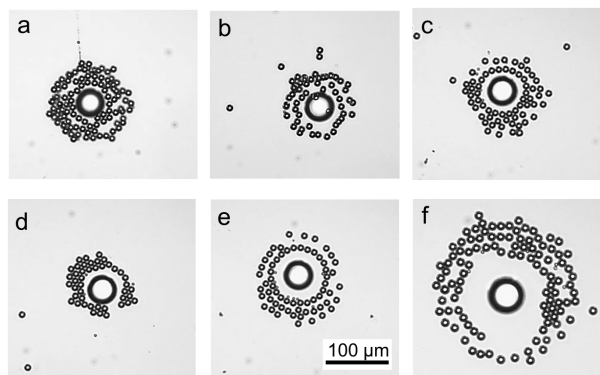

**Fig. 7** Optical micrographs of PS10 colloid assembly in eo-pumping with fixed IEX45 particles. Top row: deionized conditions; bottom row: $c_S$ = 5 μmol/L. Scale as indicated in e). a) and d) uncoated substrate; b) and e) DMOAP-1 substrate; c) and f) DMOAP-3 substrate.

Finally, we quantify the dependence of the assembly distance on the size of cargo. This was done using IEX splinters to explore their larger exchange activity yielding a long range "attraction" of colloids. DMOAP-3 and $c_S$ = 5 μmol/L were used in these experiments. Due to the slightly irregular shape of splinters, $d_S$ is not a good measure of the initial assembly distance. Instead, we measured center-center distances of the colloids and the roughly circular IEX splinters at the time of ring formation. Due to the fairly circular shape of the first line of colloids, $d_{CC}$ shows a much lower scatter than $d_S$. Moreover, for any given splinter, it appeared to be reproducible after exchanging the colloid particles by externally applied horizontal flow. Data shown in Fig. 8 are averages over some 120-150 colloid particles of PS4 and PS10 assembled at different splinters of average diameters of (40-50) μm. We also show data for PS2, which does not form a ring. Rather a void forms with a fuzzy inner boundary. This is attributed to the larger diffusivity of PS2. Here, we measured the area density $\rho(r)$ of PS2 determined from the radially averaged values of $\rho(r)$ as the distance of maximum slope $d\rho(r)/dr$. The statistical error due to the fuzziness is in this case further enlarged due to thermal drift of the solvent, which easily distorts the void shape for these light particles. The size range of colloid particles which can be used for assembly is limited. For relatively small sized cargo (2a < 500 nm), its dynamics is dominated by Brownian motion at a decreased Stokes friction, thus no ring could form. While rather big sized colloids (2a > 50 μm) settle onto the substrate, and the eo-flow friction is not able to drag it towards IEX. In the double logarithmic plot in Fig. 8, the data approximately show the $1/a^\alpha$ dependence expected from Eqn. (4). However, due to the narrow size range and few data points, the predicted power-law exponent of assembly distance with the cargo size cannot be validated satisfactorily here.

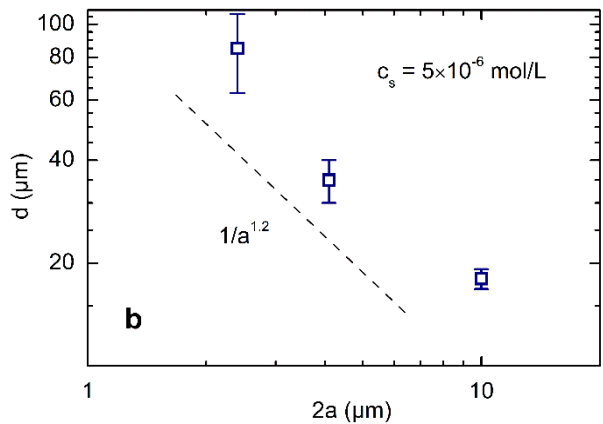

**FIG. 8 (color online)** Double logarithmic plot of the assembly distance $d$ at the time of ring formation between compact IEX splinters and colloids as a function of colloid size as determined on DMOAP-3, $c_S$ = 5 μmol/L. One observes a decrease of $d$ with colloid size. For comparison we show the theoretical expectation of $\alpha$ = 1.2 as straight dashed line.

In terms of crystal orientation, preliminary experiments using elongated IEX splinters (aspect ratio ~6) or two nearby IEX show that the crystal lattice becomes oriented along the long axis of the IEX under low flow condition (DMOAP-3 and $c_S$= 5 μmol/L, Fig. S2).

However, due to the large parameter space used in this work, we will report the orientation control in anther work.

**Simulations**

To further support the experimental observations on crystal formation, we performed BD computer simulations of purely repulsive colloidal point particles in two dimensions. Here, the time scale is set by $\tau = 1/(\kappa^2 D_0)$. The energy scale is given by $k_B T$, and the simulated unit length scale is set by the screening length $\kappa^{-1}$. In our simulations, we consider systems of $N = (20\text{-}1000)$ particles confined to disks of radii $R = (500\text{-}3500)\,\kappa^{-1}$. The chosen numbers facilitate crystal formation in reasonable simulation times. At large densities, we initially exclude a circular area of radius $R_{init} = 200\kappa^{-1}$ during the random placing of cargo particles into the disc at $t/\tau = 0$ in order to minimize possible local inhomogeneities in the initial spatial distribution of particles. After placing the particles, this constraint is removed and colloids can roam the complete space freely.

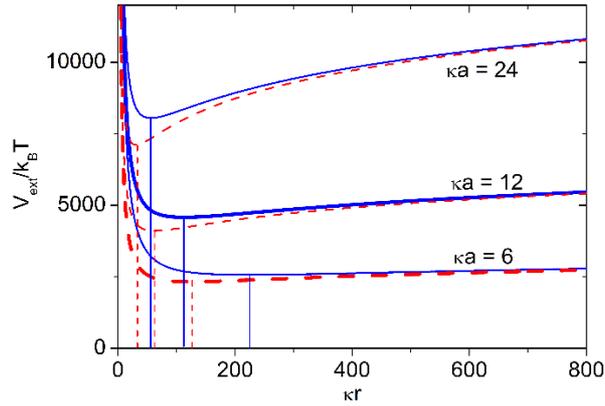

**Fig. 9 (color online)** Radial trap potentials of colloidal particles with different radii (from top to bottom): $\kappa a = 24$, $\kappa a = 12$ and $\kappa a = 6$. The dashed lines are obtained at $V_0\kappa/k_B T = 5\times10^7$, $A/D_0 = 400$, $B\kappa/k_B T = 5\times10^4$ (case 1), and the solid lines are obtained at $V_0\kappa/k_B T = 5\times10^7$, $A/D_0 = 800$, $B\kappa/k_B T = 1.8\times10^5$ (case 2). For comparison, all of the $V_{ext}/k_B T$ of case 2 are divided by a factor of 2. The vertical dashed and solid lines indicate the minima of the corresponding potentials. The thicker lines (case 1 for $\kappa a = 6$ and case 2 for $\kappa a = 12$) are the potentials used for simulations shown in Figs. 10 and 11.

The flows present in the experiments translate to two radially symmetric forces as given by Eqns. (5) and (6) in the simulations. Thus we define the corresponding radial trap potential as $V_{ext}(r) = \gamma A \ln r + B/r$. This trap potential depends on particle size through its $\kappa a$ dependence. This is illustrated in Fig. 9 for three different $\kappa a$. In addition, we compare two characteristic realizations for different reduced external field strengths. In case 1 (red dashed lines), the potential parameters are set to $V_0\kappa/k_B T = 5\times10^7$, $A/D_0 = 400$, $B\kappa/k_B T = 5\times10^4$. In case 2 (blue solid lines), $V_0\kappa/k_B T = 5\times10^7$, $A/D_0 = 800$, $B\kappa/k_B T = 1.8\times10^5$. Note that for comparison, the absolute values of case 2 are divided by a factor of 2. At each $\kappa a$, the minima of case 2 are deeper but much shallower than for case 1. A sufficiently shallow potential minimum allows colloids to assemble according to their own interactions without the need to conform to an arbitrarily shaped seed surface. Upon increasing $\kappa a$, we obtain a deeper potential and an increase in the relative steepness of the potential minimum. Therefore, at fixed reduced interaction amplitudes $V_0\kappa/k_BT$, larger particles are confined stronger in the radial direction, consequently, their thermal fluctuations become less significant for the crystallization process.

The outcome of the two representative cases is shown in simulation snapshots displayed in Figs. 10 and 11. Fig. 10 corresponds to case 1. We here started with $N = 1000$ and a disk size of $R = 500\kappa^{-1}$, the initial density of colloids is on the order of 1.3 x $10^{-3}$ $\kappa^{-2}$, and the time step was chosen as $\delta t = 2\times 10^{-5}\tau$ (Fig. 10a). We further chose $a = 6.0\kappa^{-1}$ for the colloids (small filled circles). The IEX is realized by placing a circular area of suitable extension in the disk center, which acts for the particles as hard core repulsive area (larger central circle). In Fig. 10, we chose an IEX radius of $a_{IEX} = 20\kappa^{-1}$. Crystalline particles as identified by a combination of bond order parameter ($p_6 \geq 0.85$) and bond-length-fluctuation order parameters ($b_6 \leq 0.1$) are marked in blue [54].

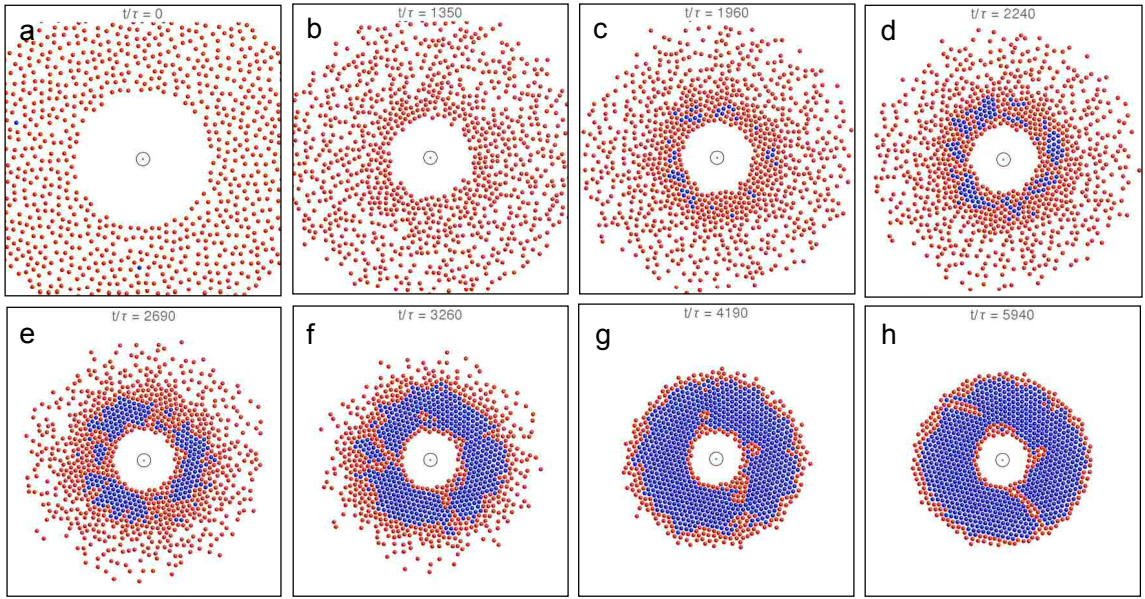

**Fig. 10 (color online)** Simulation snapshots showing the assembly of $N = 1000$ colloid particles of radius $a = 6\kappa^{-1}$ around a spherical IEX particle with size $a_{IEX} = 20\kappa^{-1}$ (circle at center) taken at the indicated times $t/\tau$ of 0, 1350, 1960, 2240, 2690, 3260, 4190, and 5940. Blue colloids show crystalline order. Field of view is 400 × 400 $\kappa^{-2}$. Simulation parameters: $N = 1000$, $\kappa R = 500$, $\kappa R_{init} = 200$, initial density 1.3 x $10^{-3}$ $\kappa^{-2}$, $\delta t = 2\times 10^{-5}\tau$ and case 1.

For case 1, we found colloid assembly to proceed qualitatively similar to the experiments with moderate flow. Figs. 10a-f show a series of characteristic simulation snapshots capturing both the initial assembly and the nearly single domain crystal formation as a function of the reduced time $t/\tau$. In this example, the accumulation of particles starts to occur at a distance $d^* = d \approx 125\kappa^{-1}$, i.e. where the net force on a single particle stemming from the forces of opposite signs is zero (c. f. Fig. 10b). At the larger particle numbers, useful for crystal growth in acceptable simulation time, no well defined ring is formed. Rather an extended concentric region of increased colloid density emerges some distance off the IEX particle. In this region, the density of colloids increases with time. Particles with high local bond-order appear within this belt (coloured blue) and crystallization begins with significant fluctuations (c.f. Figs. 10c-e and video 2). Several different crystalline grains are formed and stabilized. They intersect creating grain boundaries (Fig. 10d) that strongly fluctuate and anneal with time (Figs. 10e to 10g). At the end of this run we have obtained a bent single crystal domain with two defect zones, which either manifest themselves as grain boundaries or as dislocation lines. As compared to the low flow experiments (Fig. 3), the initial ring formation stage is absent and the final faceting is much less pronounced. This is mainly attributed to the fast assembly of colloids at large particle density. In fact, the final crystal of Fig. 9h resembles crystals grown on uncoated substrates under moderate flow conditions.

Ring formation can be observed, however, for case 2 simulations and using smaller colloid densities. An example is given in Fig. 11 for $N = 500$ colloids of size $a = 12\kappa^{-1}$ filled into a disc of $R = 3500\kappa^{-1}$ at case 2 and assembling around an IEX with size $a_{IEX} = 50\kappa^{-1}$ (see video 3 in the SI). Note that now the minimum assembly distance shows appreciable fluctuations on short time scales (c. f. Figs. 11c and 11d). Moreover, the assembly distance decreases significantly between $t/\tau = 300$ and $t/\tau = 1500$, but it becomes approximately constant thereafter. Crystal nucleation occurs again at several locations and now is correlated to facet formation which is more pronounced than in Fig. 10. Note that in this run faceting is also visible at the outer crystal boundary. The final state is composed of two crystal orientations forming from a single domain crystal with pronounced elastic deformation. Thus, depending on the adjusted parameters, the simulation captures the growth of multi-domain crystals with rough inner boundaries from assemblies without pronounced initial ring formation (video 2 in the SI) and the growth of faceted crystals after initial ring formation (video 3 in the SI).

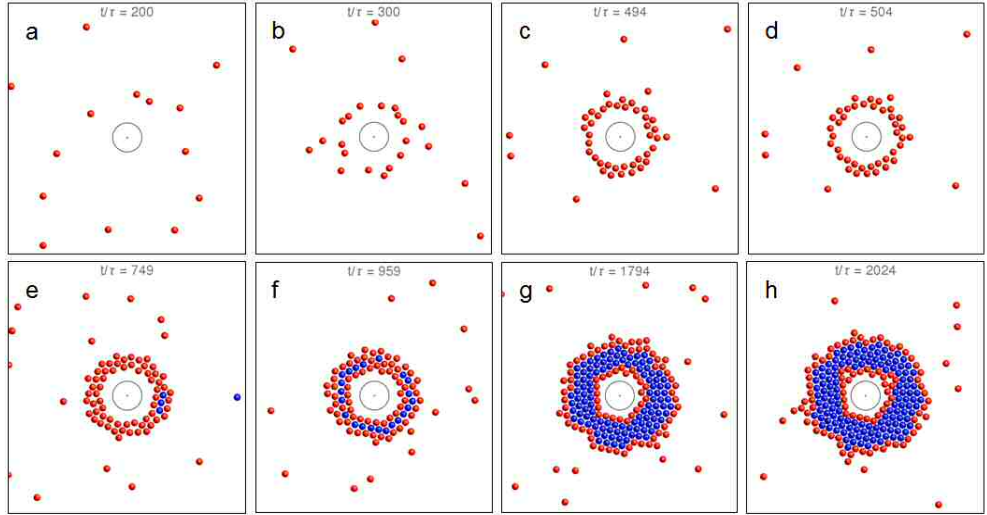

**Fig. 11 (color online)** Simulation snapshots showing the assembly of $N = 500$ colloidal particles of size a = $12\kappa^{-1}$ around a spherical IEX with size $a_{IEX} = 50\kappa^{-1}$ (central circle) taken at the indicated times of $t/\tau$ = 200, 300, 494, 504, 749, 959, 1794 and 2024. Blue particles show the crystalline order. Field of view is 400 × 400 $\kappa^{-2}$. Simulation parameters: $N = 500$, $R = 3500\kappa^{-1}$, initial density $1.3\times10^{-5}$ $\kappa^{-2}$, $\delta t = 1\times10^{-4}\tau$ and case 2.

Due to the large thermal fluctuations at the stage of ring formation and the continuous shrinking of the assembly distance at large and at low cargo density, we were not able to observe a stable initial distance of ring formation. Therefore, we measured the minimum assembly distance after equilibration of the formed crystals. In Fig. 12a, we plot the dependence of $\kappa d_{CC}$ on the particle size for the simulation case 1 at two different particle numbers of $N = 100$ and $N = 1000$. In both cases the data can be reasonably well described by a power law decrease of the assembly distance with increasing scaled particle size as can be seen from the fits (dashed lines). For the lower and larger cargo number we obtain $d \propto 1/a^{1.09}$ and $d \propto 1/a^{1.41}$, respectively. The theoretical expectation of $1/a$ is shown for comparison as dotted line. In Fig 12b, we show three sets of simulations with $N = \{20, 100, 500\}$ for case 2. All three were from runs with clearly discernible ring formation. At small $\kappa a$, all three curves coincide and show a power law decrease which compares well to the analytically expected $1/a$ behaviour. The curve for $N = 20$ shows this behaviour even up to large $\kappa a$. A stronger decrease resembling a power law with exponent larger than unity becomes visible when the assembly distance approaches the IEX radius. Qualitatively the same behaviour is seen at larger $N$, and with increasing $N$ the curves bend down to the minimum assembly distance earlier.

The decrease of assembly distance with increasing colloid size can be related to the size-dependence of the attractive trap force (Fig. 9). The assembly distance $d$ is at the minima of the external potentials, which shifts to smaller radial distance for larger $\kappa a$. A natural upper limit is given by the condition that the minimum should lie outside the IEX. A natural lower limit is given by the thermal fluctuations which for $\kappa a < 3$ become exceedingly large, thus defying measurements of $d$. The prescribed radius dependence is however fulfilled strictly only for the case of a single colloid. Once a second or third layer forms, we observe a decrease of $d$ in time and an obvious deviation of the final assembly distance from scaling theory. Thus the curves for larger particle density lie below those at low density. For case 2, they in addition show a qualitative deviation from power law scaling at large $\kappa a$ when the position of the minima approaches the IEX radius (horizontal dashed lines in Figs. 12a and 12b).

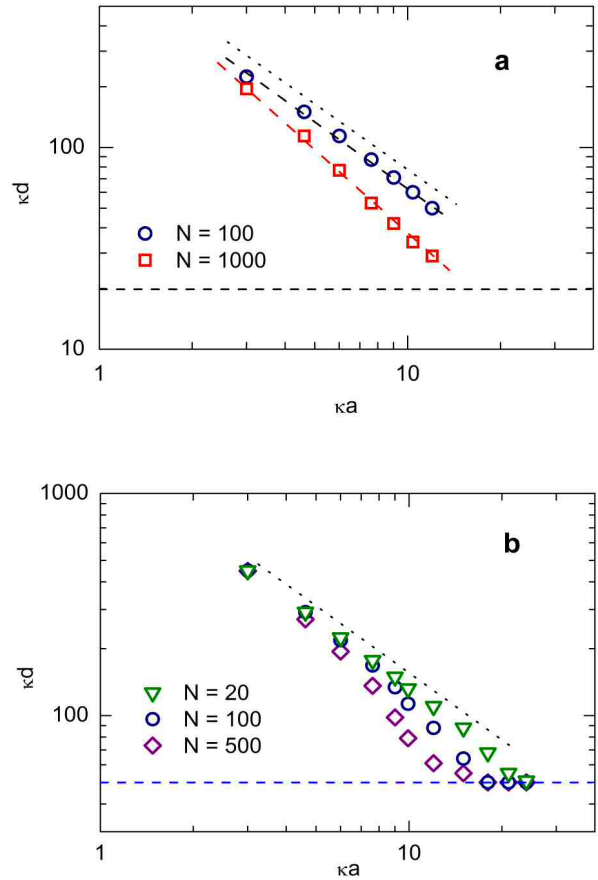

**Fig. 12 (color online)** The reduced minimum separation distance $\kappa d$ as a function of the reduced particle size $\kappa a$. a) Case 1, $N = 100$ (circles) and $N = 1000$ (squares). The dashed lines are power law fits yielding $d \propto 1/a^{1.09}$ and $d \propto 1/a^{1.41}$, for $N = 100$ and $N = 1000$, respectively. b) Case 2 and $N = 20$ (triangles), $N = 100$ (circles) and $N = 500$ (diamonds). The horizontal dashed lines at the bottom of both plots mark the IEX size, i.e. the geometrically possible minimum assembly distance. In both plots, the dotted line of slope -1 gives the analytical expectation for the assembly distance of an individual colloidal sphere, i.e. in the absence of colloidal interactions.

**Discussion**

We have assembled single crystals at pre-determined positions using an ion-exchange-based electro-osmotic pumping. We systematically changed the experimental boundary conditions, e.g., the exchange rate of IEX, mobility of substrate, and mobility and size of colloid. The essential finding is that a finite assembly distance is crucial for the formation of high quality crystals (c.f. Fig. 5), because it allows for sufficient freedom of rearrangement and does not force the colloidal arrangement to copy the contours of the IEX particle. The analytic theory predicts the dependence of assembly distance on colloid size and mobility (included in $Z_{eff}$) and substrate mobility ($\mu_{eo}$). The size dependence of assembly distance is also tested by simulation which further shows that the decrease of assembly distance with the size of colloid is set for the initial state (c.f. Fig. 9) and is also found in the final state of assembly (c.f. Fig. 12). We stress that a strong requirement for the scaling theory is low colloid density, because at high density colloid-colloid interaction decreases the assembly distance and thus leads to the breakdown of the power law scaling. Furthermore, a low colloid density is essential for assembly high quality crystals, as strong colloid-colloid repulsion at early times enhances the number of formed nuclei and suppresses their annealing quite effectively (c.f. Fig. 10). However, our analytic theory and simulation are limited to two-dimensional (low) flow condition. Once upward convection intervenes in the near field at moderate flow, or three-dimensional flow appears at tall cell geometry [40], simple modelling is not feasible any more.

In terms of extension and outlook to the present approach, one could use line shaped IEX arrangement to control the orientation of crystal lattice, as indicated by Fig. S2. This also opens up a new way of creating interesting crystal microstructures if the seed with a designed structure can be fixed on the substrate, e.g, using capillary force [55]. The microstructure design is under work, and the results will be presented in another paper. More importantly, the approach used in this work could be implemented to other types of pumps (e.g., thermo-phoretic pump or catalytic pump [35,56]) under the condition that two opposing flows compete with each other. Then of course, one could also introduce an on-off switch function by utilizing pump with external trigger (e.g. UV-illuminated AgCl [28], light induced diffusioosmosis [57]). Consequently, the assembly can be terminated at any stage, allowing for interesting kinetic experiments. This would provide a facile and controlled extension of the pioneering experiments on free expansion melting [58]. Finally, the issue of fixing the single crystal structures could be achieved using photo-polymerization of charge neutral monomers [59].

**Conclusions**

We assembled individual colloidal single crystals at pre-defined locations using eo-pumps based on ion exchange at μmolar levels of impurity concentration. We gave the first experimental characterization and a simple scaling theoretical modelling of the processes involved in assembly under low flow conditions, namely a competition of two opposing electro-kinetic flows with different radial dependence of their magnitudes. This approach was further implemented in our BD simulations which reproduced the kinetics of the assembly process remarkably well. In addition, they turned out indispensable for clarifying the cause of some of the experimental observations. From our results we now understand that the obtainable crystal quality is primarily resulting from slow initial assembly at a finite distance off the IEX surface. The focus of the present paper was on the understanding of crystal assembly at ion exchange based micro-pumps. However, our results may also contribute to a deepened understanding of the electric contributions in other micro-pumps based on chemical reactions or heat production. Finally, we shortly touched some fascinating potential applications that may be enabled by further development of the technical implementation.

**Acknowledgement**

It is a pleasure to thank Joost de Graaf and Christian Holm for intense discussions on the microphysics of diffusio-phoresis and eo-pumping. We gratefully acknowledge the DFG for financial support (SPP 1296 and SPP1276, Grant Nos. Pa459/17, Pa459/18, Lo418/15, and Lo418/17). H. M. was receiving a stipend of "Sparkassenverband Rheinlad-Pfalz" during her "Jugend forscht" awarded internship at JGU.

# Controlled assembly of single colloidal crystals using electro-osmotic micro-pumps

Ran Niu,[a],† Erdal C. Oğuz,[b,c] Hannah Müller,[a] Alexander Reinmüller,[a] Denis Botin,[a] Hartmut Löwen[b] and Thomas Palberg[a]

a. Institute of Physics, exp. Soft Matter Group KOMET336, Johannes Gutenberg University, D-55099 Mainz, Germany.
b. Institute for Theoretical Physics II: Soft Matter, Heinrich-Heine-University, D-40225 Düsseldorf, Germany.
c. present address: School of Mechanical Engineering and The Sackler Center for Computational Molecular and Materials Science, Tel Aviv University, Tel Aviv 69978, Israel.
† Corresponding author: Ran Niu, Email: ranniu@uni-mainz.de

**Experimental**

To qualitatively check the exchange rate per mass, we performed measurements of the decreasing conductivity in standard electrolyte solutions using a standardized closed deionization circuit,[S1] each with 10 g of different species of IEX added. We found the exchange rate per mass of the resin based IEX to be about one order of magnitude larger than for the micro-gel IEX. For the later we further observed a decrease of the rate with decreasing size. We also quantified the exchange rate of a single fixed IEX45 by measuring the radial distribution of pH-values as a function of time in a typical experimental geometry. Comparing the concentration of released protons integrated over the formed gradient at different times, we obtain an exchange rate of $5 \times 10^{-17}$ mol/s, decreasing with $t^{-1/2}$, as expected for diffusion-limited exchange.

The electrophoretic mobility of isolated colloidal spheres was measured in deionized water at contact with ambient air by micro-electrophoresis in a home-build Perspex cell (10 mm×10 mm) based on the construction originally introduced by Uzgiris.[S2] An alternating square wave electric field ($f$ = 0.5 Hz; $E$ = 20 V/cm) was applied between two platinum electrodes mounted vertically in the cell center far off the cell walls to avoid influence of electro-osmosis. The cell was mounted on a micro-electrophoresis instrument (Mark II, Rank Bros. Bottisham, Cambridge, UK) providing ultra-microscopic illumination and observed with a consumer DSLR (D800, Nikon, Japan) equipped with macro-lens. Using an exposure time of 3 s, the response of individual spheres to gravity and the field translated to zigzagged trajectories readily analysed for the electro-phoretic mobility. Data were averaged over at least $5 \times 10^2$ particles for each species. We used standard electro-kinetic theory,[S3] to calculate the effective charges $Z_{eff}$ assuming a Debye-Hückel-type electro-static potential (see Eqns. 9a and 9b in the main text). Results for $Z_{eff}$ as a function of particle diameter are shown in Fig. S1. Over the range of diameters $Z_{eff}$ shows a power-law dependence and the best least square fit yields $Z_{eff} \propto a^{1.8}$.

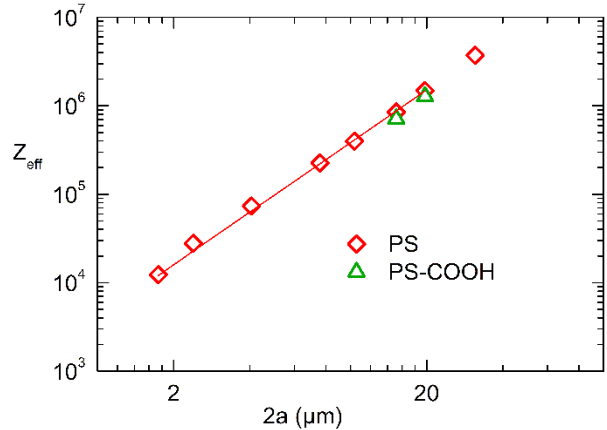

**Fig. S1** Effective cargo charge determined from micro-electrophoresis versus colloid diameter. Data for two different colloid species of differing surface chemistry are shown.

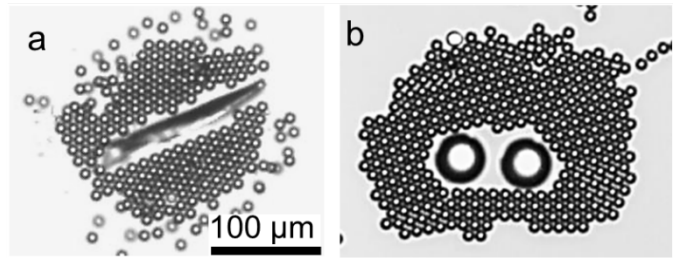

**Fig. S2** Optical micrographs of orientated crystal formed on DMOAP-3, $c_S$ = 5 μmol/L by PS10: a) at an elongated IEX splinter, b) at two nearby IEX45. 100 μm scale bar applies for both images. The crystal lattice becomes oriented along the long axis of the IEX.

**Videos:**

Video 1: PS5 circulating in the anisotropic convection cell of an IEX splinter on uncoated substrate. (Image size: 224x168 μm$^2$, 20x real time speed).

Video 2: The assembly of colloidal particles of radius $a = 6\kappa^{-1}$ around a spherical IEX particle with size $a_{IEX} = 20\kappa^{-1}$. Field of view is 400 × 400 $\kappa^{-2}$. Simulation parameters: $N = 1000$, $\kappa R = 500$, $\kappa R_{init} = 200$, initial density 1.3 x 10$^{-3}$ $\kappa^{-2}$, $\delta t = 2\times10^{-5}\tau$ and case 1. At larger particle density, no well defined ring is formed. Crystallization begins with significant fluctuations. Several different crystalline grains are formed and stabilized, which intersect creating grain boundaries that strongly fluctuate and anneal with time. At the end of this run, a bent single crystal domain with two defect zones is obtained.

Video 3: The assembly of colloidal particles of radius $a = 12\kappa^{-1}$ around a spherical IEX particle with size $a_{IEX} = 50\kappa^{-1}$. Field of view is 400 × 400 $\kappa^{-2}$. Simulation parameters: $N = 500$, $R = 3500\kappa^{-1}$, initial density $1.3\times10^{-5}$ $\kappa^{-2}$, $\delta t = 1\times10^{-4}\tau$ and case 2. At smaller colloid density, ring formation is observed, and the assembly distance decreases significantly between $t/\tau = 300$ and $t/\tau = 1500$, but it becomes approximately constant thereafter. In the final state, a single domain crystal with two crystal orientations and pronounced elastic deformation forms.